# Title page

**Author affiliations:**

B. Fichtl[1,2], M. F. Schneider[3]

[1]University of Augsburg, Experimental Physics I, Augsburg, 86159, Germany.

[2]Nanosystems Initiative Munich NIM, Schellingstr. 4, 80799 München, Germany.

[3]Medizinische und Biologische Physik, Technische Universität Dortmund, Otto-Hahn Str. 4, 44227 Dortmund, Germany.

**Corresponding author:**

Matthias F. Schneider

| | |
|---|---|
| Address: | Medizinische und Biologische Physik, Technische Universität Dortmund, Otto-Hahn Str. 4, 44227 Dortmund, Germany |
| Telephone: | (+49) 231 755 4139 |
| E-Mail: | matschnei@googlemail.com |

# Abstract

Biological signaling is imagined as a combination of activation and transport. The former is triggered by local molecular interactions and the latter is the result of molecular diffusion. However, other fundamental physical principles of communication have yet to be addressed. We have recently shown, that lipid interfaces allow for the excitation and propagation of sound pulses. Here we demonstrate, that these reversible perturbations can control the activity of membrane embedded enzymes without the necessity of molecular transport. They therefore allow for the rapid communication between distant biological entities (e.g. receptor and enzyme) at the speed of sound, which is here in the order of 1 m/s within the membrane. The mechanism reported provides a new physical framework for biological signaling.

# Text

How cells communicate and interact and how all the countless localized processes within a cell are integrated are two of the most central questions of cell biology. Currently cell signaling is described as a cascade of binding and molecular activation events with diffusion as the only physical mechanism for transport. A famous example, awarded with the Nobel Prize for Physiology in 1994 and recently for chemistry 2012, is the extracellular binding of a hormone to a G-Protein Coupled Receptor. This subsequently leads to the intracellular release of molecules and thus to the activation of specific proteins (e.g. kinases). They, in turn, liberate further molecules (2nd messengers, cAMP), which eventually fulfill some biological function (e.g. the catalysis of Glycogen into Glucose) (1). In principle, an extracellular stimulus was transmitted to the inside of the cell, where the signal led to a particular biological response (e.g. Glucose production). The underlying physical principle is the binding induced change of a proteins conformation, possibly leading to catalytic steps or disintegration of protein complexes, followed by the diffusion of released molecules (subunits, catalytic products etc.) (2). We suggest a completely new mechanism based on propagating pulses.

The continuous membrane constantly experiences all sorts of external perturbations. Pressure and temperature variations, changes in ionic conditions or pH (as induced during many catalytic reactions) as well as the binding of proteins (incl. second messengers) to name only a few sources. If the membrane is sufficiently decoupled from the bulk over the timescale of the perturbation, the perturbation will begin to propagate along the membrane. Importantly, this is not a hypothesis, but inevitable and follows directly from the second law of thermodynamics. (Footnote: If propagation is impaired but dissipation still suppressed, a perturbation may lead to local oscillations, which we don't discuss here as we consider the membrane as a continuous system). This idea dates back to Konrad Kaufmann, who proposed in the late 80s' to start from the second law and realize that the membrane represents an independent thermodynamic system and therefore is expected to be excitable for propagating sound pulses (3–5).

In a recent series of experiments, we were indeed able to demonstrate, that lipid interfaces are capable to efficiently support the propagation of linear and non-linear acoustic pulses (6–12). The decoupling probably results from the large difference in compressibility between membrane and water, which implies a poor impedance coupling similar to the principal of wave guiding (6). Type of wave and excitation span a wide range. Optical, mechanical as well as chemical (pH) excitation has been demonstrated (7, 12, 13). Interestingly, these variables not only allow to evoke pulses, but all of them are directly affected by the propagating pulses, too: the local pH at the monolayer changes in phase with fluorescence emission, surface potential and surface pressure (8, 9, 12). Hence the interfacial pulses correspond to a reversible thermodynamic state change of *all* thermodynamic variables, which opens up entirely new possibilities for cell communication. Needless to say communication by sound "beats" diffusion clearly in all respects: it is adiabatic, directed, easily faster, scales linear with distance (and thus works well not only over small but also over large distances) and finally does not require vast amounts of material to be transported in order to communicate. The experiments presented here provide clear evidence how pulses, propagating within a lipid membrane, regulate the catalytic activity of distant, embedded enzymes. They provide therefore strong support for the role of adiabatic propagations in biological communication and signaling.

Our system – a simple mimic of a cell membrane - is based on an enzyme embedded in an excitable lipid monolayer. The setup allows to locally detect the mechanical component of the two dimensional propagating pulse simultaneously with enzymatic activity and hence gives direct access to pulse-enzyme interactions (Fig. 1). As recently demonstrated by us a "puff" of acidic gas leads to the excitation of a propagating pulse (12). Stimulating the monolayer with hydrogen chloride, acetic acid as well as carbon dioxide indicates the independence of the excitation process from the respective Brønsted acid. Furthermore the excitation strength strongly depends on the $pK_a$ of the lipid monolayer, illustrating the origin of the phenomenon: the protonation of the lipid head groups due to local acidification. The velocity of the pulses is determined by the state of the monolayer, represented by a certain compressibility. During the phase transition the compressibility increases strongly, i.e. the monolayer is getting softer, which leads to a minimum in propagation speed. This was reported in more detail earlier and demonstrates that the pulse propagation is indeed an adiabatic effect (7).

In order to test the hypothesis, that these pulses are capable of forming a new pillar in biological communication, we incorporated the enzyme Acetylcholinesterase (AChE), extracted from Torpedo californica, into the lipid monolayer. The dimeric form bears a covalently attached glycophosphatidylinositol (GPI) anchor, which was posttranslationally added at the C terminus of each subunit (14–16). The overall four fatty acid chains are responsible for anchoring the enzyme to the membrane surface (17). The enzymatic activity is measured by the cleavage of the chromophoric substrate acetylthiocholine at a wavelength of around 410 nm (18, 19).

Figure 2A and B show the simultaneous detection of enzyme activity and lateral pressure during a travelling pulse following an excitation by HCl. The pulse propagates along the surface at a speed of ~0.5 m/s. Upon arrival at the window of detection, where the enzymatic activity is recorded, substrate hydrolysis and hence the catalytic rate is clearly altered (Fig. 2A). After a strong (~ 19 fold) increase, the enzymatic activity drops to zero (Fig. 2B). The biphasic activity of the enzyme correlates well with the lateral pressure signal of the expanding and condensing monolayer. We would like to note that this biphasic behavior was consistently observed, even though the first phase of increased activity varies from measurement to measurement (for more examples see S2).

Higher time resolution and more details can be resolved when excitation is induced by $CO_2$ instead of HCl. In water $CO_2$ partly reacts to $HCO_3^-$ and $H^+$ on a very slow timescale ($k_{H^+}$ ~ 0.0375 1/s) (20). Consequently, the resulting pulses are much broader than in the case of HCl (see Fig. 2C/D). Nevertheless, the biphasic activity pattern is conserved. During the expanding wave front the activity increases ~2 fold, while during the subsequent condensation back to equilibrium substrate cleavage almost stops completely.

To assure we indeed observe the impact of the propagating pulse on the enzymatic activity proper control experiments have been performed in the absence of enzyme, but presence of everything else (substrate, assay components, lipids etc.). No effect of the pulses on the substrate as well as on the reaction product were detected (Fig. S3A). Importantly, these quasi-adiabatic results are in stark contrast to the behavior of the enzymes during isothermal measurements: in this case the catalytic rate of the enzyme in the respective pressure range of the monolayer is approximately constant (Fig. S3B).

As argued before (6, 12), the pulses we report on are acoustic in nature and must not be mistaken with chemical waves in excitable materials reported earlier (21–23), where diffusion is the underlying mechanism for transport. Here, no material transport is required for the propagation and communication of a signal. Further, in principle, no energy is required for transport either given the adiabatic nature of the pulse. Thermodynamically, the stimulated pulse can be seen as an (adiabatic) state change propagating along the interface. Consequently, an enzyme located within the acoustical path of the pulse experiences a transient change of all of its thermodynamic variables. Degree of state change depends on the amplitude of the propagating pulse as well as the initial conditions of the membrane patch embedding the enzyme. The fact that the membrane environment can regulate the activity of enzymes is well known from quasi-static experiments (24–28). While in our case isothermal changes of the lipid monolayer in the liquid-expanded phase exert only a minor influence on enzyme activity (cf. S3B), the effect of the adiabatic pulses is striking.

As mentioned above, all observables of the system have to be coupled (cf. Maxwell relations) (29). For instance, when a charged membrane is compressed, not only pressure and temperature increase reversible, but also the charge density. Hence, along with a pressure pulse we have to expect a propagating change in surface potential, which we have indeed directly and simultaneously observed earlier (8, 12). Consequently, ion distribution and in particular (the very mobile) protons will rearrange under the presence of the propagating electric field, which leads to the propagation of a pH pulse simultaneously with 2D-pressure, voltage and temperature. The existence of a propagating pH pulse was indeed confirmed recently (12). Reversible pH changes at the interface of up to +/-1 unit can be easily realized. A pressure drop in this lipid system is associated with a local increase in pH (12), which would indeed give rise to an increase in AChE activity. The latter follows from the pH dependence of AChE which experiences a pH-optimum at a basic bulk pH of ~8 (30). According to Silman (see Fig. 5) a pH increase from 6.5 to 7.5 (as observed during pulse propagation), would give rise to activity changes of ~2 in excellent agreement with $CO_2$ excitation (cf. Fig. 3). When, however, excited by HCl timescales shorten and activity changes increase. Although, it is beyond the scope of this paper to discuss these results quantitatively a few potential sources of this discrepancy should be mentioned: pulses are not isothermal but adiabatic. Since the HCl-excitation leads to very fast local changes of the interfacial variables, the process resembles much more an ideal adiabatic state change than the relatively slow $CO_2$ induced changes, which fall somewhere in between the two extreme cases (adiabatic vs. isothermal). Taking the proton perspective – since protons are part of the reaction (they are released by the cleavage of acetylthiocholine) - the reaction equilibrium is shifted differently. In the quasi-adiabatic case the rapid change results in a transient release of protons accompanied by a shift towards product production, which is much less pronounced in the quasi-isothermal case. Of course, interaction between pulses and enzymes is not limited to pH effects but depends on all thermodynamic variables. E. g. enzyme activity is often coupled to other nonlinear physical quantities, like, for example electrical capacity or compressibility, turning this general phenomenon in a very specific process (27, 31).

One surprising observation of our results is the asymmetric biphasic change in activity for the pH induced pulse. Therefore the question arises, why activity changes do no follow state changes (pressure variations), but appear biphasic (cf. Fig. 2). In other words, why is the initial increase in activity followed by a sudden phase of no activity (a "blocked" enzyme) instead of a continuous decrease back to baseline for the pH induced pulses?

As established before, every pulse represents a propagating thermodynamic state change. Hence its interaction with proteins will strongly depend on the local and current state of the interface, the strength of the perturbation as well as on the chemo-physical properties of the single protein. Therefore we would like to restrain from giving an explicit molecular picture. Nevertheless we hypothesize, that the biphasic activity of the AChE in the case of the pH induced pulses might be caused by the following effect: In the first phase (1) the pulse-induced pH increase leads to an increase in enzyme activity (see Fig. 3). Due to the strongly enhanced degradation of substrate, the local proton concentration around the enzyme drastically increases. Combined with the back diffusion of the liberated protons from the lipids, this might lead to a temporary pH drop at the interface and hence to a decrease in enzyme activity (2). Finally, the system relaxes back to initial conditions as the perturbation disappears (3).

It would be thrilling to follow these catalytic changes on a single molecular level, which could be achieved by FCS. Kaufmann has developed a theory based on the second law of thermodynamics, according to which changes in catalytic activity represent changes in fluctuation strength of the reaction coordinate induced by the absorption of the substrate. This theory could be tested with our system and the tools of FCS.

As a concluding remark, we would like to mention that in order to study whether the observed pulse-enzyme interaction is specific for AChE, we performed a series of experiments with the enzyme phospholipase A2. While further experiments are still running, first results show that the enzyme activity of PLA2 decreases during enzyme pulse-interaction. Therefore we propose that acoustic pulses as an essential mechanism for orchestrating different processes in cells (Fig. 4).

## Materials & Methods

**Chemicals**:

Acetylthiocholine (ATC), Ellman's reagent (5,5'-dithiobis-[2-nitrobenzoic acid]) (DTNB) as well as hydrochloric acid (32%) were purchased from Sigma-Aldrich (USA). 1,2-dimyristoyl-sn-glycero-3-phospho-L-serine (DMPS) was acquired from Avanti Polar Lipids (USA) and used without further purification. The lipids were dissolved in a chloroform/methanol/water solution at a concentration of 10 mg/ml. Detergent-soluble Acetylcholinesterase from *Torpedo californica* was kindly provided by Prof. Isreal Silman (Weizmann Institute). It was purified by affinity chromatography from electric organ tissue of Torpedo californica after solubilization in 1% sodium cholate (32).

**Experimental setup:**

For the enzyme-pulse experiments a custom-made Langmuir trough (NIMA) with an area of 150 cm$^2$ (Fig. S1) was used. The propagating mechanical component of the pulse was measured by a pressure sensor with up to 10 000 samples per second and 0.01 mN/m resolution. The sensor was situated next to the window of detection (cf. Fig. S1), where simultaneously the activity of the enzymes was determined by the chromophoric Ellman Assay. The hydrolysis of ATC leads to the release of thiocholine, which is immediately reduced by DTNB, finally yielding an intense yellow chromophore with an absorption peak at around 412 nm (18, 19). Changes in the absorbance of the light emitted by a LED located above the trough, due to generation of the chromophore, were followed using a photodiode placed underneath the window (cf. Fig. 1).

**Experimental procedure:**

In order to study the effect of the HCl-induced pulses on enzyme activity the trough was filled with ultrapure water (resistivity > 18 MΩcm), containing 100 mM NaCl, 10 mM phosphate buffer, 2 mM ATC and 2 mM DTNB. Whereas for the experiments with $CO_2$ excitation concentrations of 10 mM NaCl, 1 mM phosphate buffer, 2 mM ATC and 2 mM DTNB were used. A DMPS monolayer is generated by titrating a few microliters of the dissolved lipid onto the surface of the aqueous solution until the designated lateral pressure is achieved. After 10 minutes solvent evaporation, 3 μl of the enzyme solution (c = 68 μg/ml) are added in the left part of the trough. To assure a homogeneous enzyme distribution the subphase is gently stirred for 5 minutes. After further 15 minutes equilibration time pulse-enzyme-interaction is studied. HCl-induced pulses are evoked by blowing 25 ml $N_2$ through the gas phase of a bottle filled with 32% hydrochloric acid. For the $CO_2$-induced pulses pure $CO_2$ gas is blown for 15 s with a flux of 150 ml/min onto the lipid monolayer.

## Acknowledgments

M.F.S. thanks Dr. Konrad Kaufmann (Göttingen), who not only inspired him to work in this field, but also initiated this project. We also thank him for numerous seminars and discussions. We are very grateful to Prof. Achim Wixforth and his chair (Experimental Physics 1 – University of Augsburg) for the support of this project. Furthermore we would like to thank Prof. Israel Silman for providing the enzyme and assisting us in all Acetylcholinesterase related questions. M.F.S. appreciates funds for guest professorship from the German research foundation (DFG), SHENC-research unit FOR 1543. B. F. is grateful to Studienstiftung des deutschen Volkes and DFG (FOR 1543) for funding.

# References

1. Scott JD, Pawson T (2009) Cell signaling in space and time: where proteins come together and when they're apart. *Science* 326(5957):1220–1224.

2. Lefkowitz RJ (2012) Nobel Lecture: A Brief History of G Protein Coupled Receptors. Available at: http://www.nobelprize.org/nobel_prizes/chemistry/laureates/2012/lefkowitz-lecture.html.

3. Kaufmann K (1977) On the kinetics of acetylcholine at the synapse. *Naturwissenschaften* 64(7):371–376. Available at: http://link.springer.com/10.1007/BF00368736 [Accessed November 19, 2015].

4. Kaufmann K, Silman I (1983) The induction by protons of ion channels through lipid bilayer membranes. *Biophys Chem* 18(2):89–99. Available at: http://www.sciencedirect.com/science/article/pii/0301462283850029 [Accessed March 8, 2016].

5. Kaufmann K (1980) *Acetylcholinesterase und die physikalischen Grundlagen der Nervenerregung*. (Universität Göttingen, Göttingen).

6. Griesbauer J, Wixforth a, Schneider MF (2009) Wave propagation in lipid monolayers. *Biophys J* 97(10):2710–2716. Available at: http://www.pubmedcentral.nih.gov/articlerender.fcgi?artid=2776282&tool=pmcentrez&rendertype=abstract [Accessed August 17, 2011].

7. Griesbauer J, Bössinger S, Wixforth A, Schneider MF (2012) Propagation of 2D Pressure Pulses in Lipid Monolayers and Its Possible Implications for Biology. *Phys Rev Lett* 108(19):198103. Available at: \url{http://link.aps.org/doi/10.1103/PhysRevLett.108.198103}.

8. Griesbauer J, Bössinger S, Wixforth A, Schneider MF (2012) Simultaneously propagating voltage and pressure pulses in lipid monolayers of pork brain and synthetic lipids. *Phys Rev E* 86:61909. Available at: \url{http://link.aps.org/doi/10.1103/PhysRevE.86.061909}.

9. Shrivastava S, Schneider MF (2013) Opto-Mechanical Coupling in Interfaces under Static and Propagative Conditions and Its Biological Implications. *PLoS One* 8(7):e67524.

Available at: \url{http://dx.doi.org/10.1371/journal.pone.0067524}.

10. Shrivastava S, Schneider MF (2014) Evidence for two-dimensional solitary sound waves in a lipid controlled interface and its implications for biological signalling. *J R Soc Interface* 11(97).

11. Shrivastava S, Kang KH, Schneider MF (2015) Solitary shock waves and adiabatic phase transition in lipid interfaces and nerves. *Phys Rev E* 91(1):12715. Available at: http://link.aps.org/doi/10.1103/PhysRevE.91.012715.

12. Fichtl B, Shrivastava S, Schneider MF (2016) Protons at the speed of sound: Predicting specific biological signaling from physics. *Sci Rep* 6:22874. Available at: http://dx.doi.org/10.1038/srep22874.

13. Suzuki M, Möbius D, Ahuja R (1986) Generation and transmission of a surface pressure impulse in monolayers. *Thin Solid Films* 138(1):151–156. Available at: http://www.sciencedirect.com/science/article/pii/0040609086902257 [Accessed September 22, 2015].

14. Futerman AH, Low MG, Silman I (1983) A hydrophobic dimer of acetylcholinesterase from Torpedo californica electric organ is solubilized by phosphatidylinositol-specific phospholipase C. *Neurosci Lett* 40(1):85–89.

15. Futerman AH, Fiorini RM, Roth E, Low MG, Silman I (1985) Physicochemical behaviour and structural characteristics of membrane-bound acetylcholinesterase from Torpedo electric organ. Effect of phosphatidylinositol-specific phospholipase C. *Biochem J* 226(2):369–377.

16. Low MG, Futerman AH, Ackermann KE, Sherman WR, Silman I (1987) Removal of covalently bound inositol from Torpedo acetylcholinesterase and mammalian alkaline phosphatases by deamination with nitrous acid. Evidence for a common membrane-anchoring structure. *Biochem J* 241(2):615–619.

17. Silman I, Futerman AH (1987) Modes of attachment of acetylcholinesterase to the surface membrane. *Eur J Biochem* 170(1–2):11–22. Available at: http://www.ncbi.nlm.nih.gov/pubmed/3319614.

18. Ellman GL, Courtney KD, Andres JVR, Feather-Stone RM (1961) A new and rapid

colorimetric determination of acetylcholinesterase activity. *Biochem Pharmacol* 7:88–95.

19. Massoulié J, Pezzementi L, Bon S, Krejci E, Vallette FM (1993) Molecular and cellular biology of cholinesterases. *Prog Neurobiol* 41(1):31–91. Available at: http://www.ncbi.nlm.nih.gov/pubmed/8321908.

20. Gibbons BH, Edsall JT (1963) Rate of Hydration of Carbon Dioxide and Dehydration of Carbonic Acid at 25 Degrees. *J Biol Chem* 238:3502–3507.

21. Ross J, Muller SC, Vidal C (1988) Chemical waves. *Science* 240(4851):460–465.

22. Sager B, Kaiser D (1994) Intercellular C-signaling and the traveling waves of Myxococcus. *Genes Dev* 8(23):2793–2804. Available at: http://genesdev.cshlp.org/content/8/23/2793.abstract.

23. Loose M, Fischer-Friedrich E, Ries J, Kruse K, Schwille P (2008) Spatial regulators for bacterial cell division self-organize into surface waves in vitro. *Science* 320(5877):789–792.

24. Sandermann HJ (1978) Regulation of Membrane Enzymes By Lipids. *BBA* 515:209–237.

25. Skou JC (1959) Studies on the influence of the degree of unfolding and the orientation of the side chains on the activity of a surface-spread enzyme. *Biochim Biophys Acta* 31(1):1–10. Available at: http://linkinghub.elsevier.com/retrieve/pii/0006300259904329.

26. Verger R (1976) Interfacial enzyme kinetics of lipolysis. *Annu Rev Biophys Bioeng* 5:77–117.

27. Op den Kamp JA, Kauerz MT, van Deenen LL (1975) Action of pancreatic phospholipase A2 on phosphatidylcholine bilayers in different physical states. *Biochim Biophys Acta* 406(2):169–77.

28. Hønger T, Jørgensen K, Biltonen RL, Mouritsen OG (1996) Systematic Relationship between Phospholipase A2 Activity and Dynamic Lipid Bilayer Microheterogeneity. *Biochemistry* 35(28):9003–9006. Available at: http://dx.doi.org/10.1021/bi960866a.

29. Steppich D, et al. (2010) Thermomechanic-electrical coupling in phospholipid monolayers near the critical point. *Phys Rev E* 81(6):61123. Available at: \url{http://link.aps.org/doi/10.1103/PhysRevE.81.061123} [Accessed November 27,

2014].

30. Silman HI, Karlin A (1967) Effect of local pH changes caused by substrate hydrolysis on the activity of membrane-bound acetylcholinesterase. *Proc Natl Acad Sci U S A* 58(4):1664–1668.

31. Maggio B (1999) Modulation of phospholipase A2 by electrostatic fields and dipole potential of glycosphingolipids in monolayers. *J Lipid Res* 40(5):930–939.

32. Futerman AH, Low MG, Michaelson DM, Silman I (1985) Solubilization of membrane-bound acetylcholinesterase by a phosphatidylinositol-specific phospholipase C. *J Neurochem* 45(5):1487–1494. Available at: http://www.ncbi.nlm.nih.gov/pubmed/4045459.

# Figures & Figure Legends

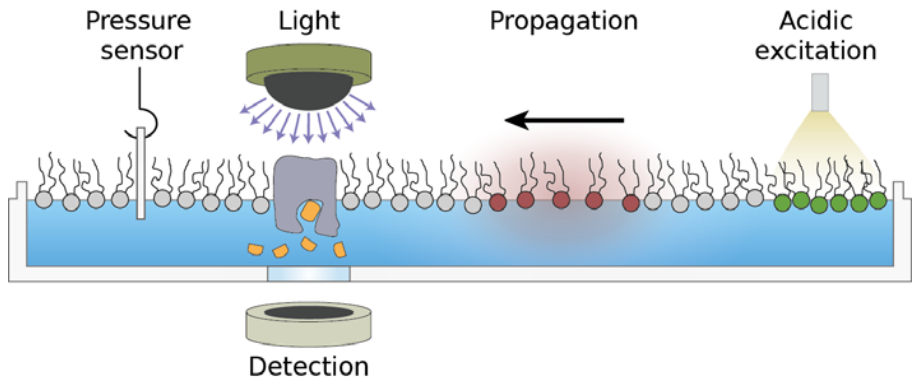

**Fig. 1**. Langmuir trough setup for measuring pulse-enzyme interaction: pulses are excited on a lipid monolayer via locally adding gaseous acid (HCl or $CO_2$). The protonation of the lipid head groups evokes a propagating pulse whose mechanical component is recorded by a pressure sensor. Simultaneously, the activity of the enzyme Acetylcholinesterase is monitored by the colorimetric Ellman Assay, which is based on the absorption of the reaction product at ~ 410 nm.

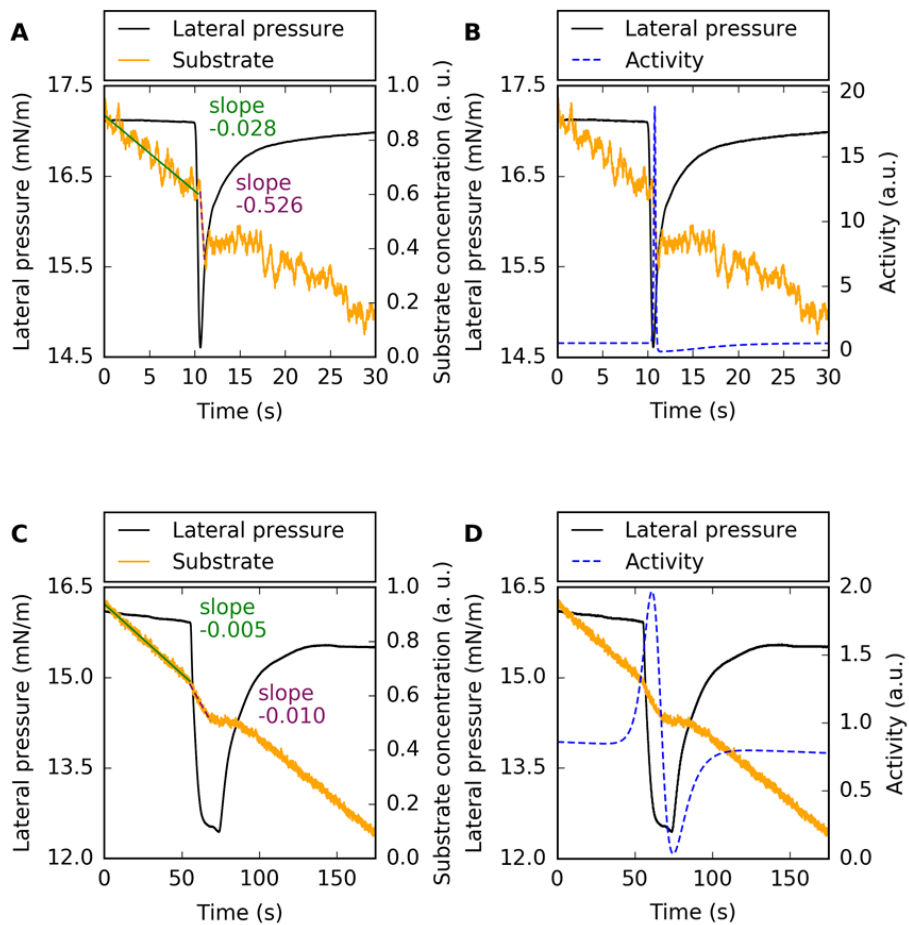

**Fig. 2.** Enzyme-pulse interaction: **(A)** The addition of hydrogen chloride onto a lipid monolayer locally excites a lateral pressure pulse. The resulting perturbation propagates with a velocity of ~0.5 m/s across the monolayer. The enzyme AChE is part of the interface and responds to the pulse with a change of its catalytic rate, which is visible by the nonlinear change in substrate concentration. The slope of the curve corresponds to the activity of the enzyme, which increases ~19fold during the first (expanding) phase of the pulse (slope 1 = -0.028, slope 2 = -0.526, factor = 18.8). Whereas during the subsequent condensation back to equilibrium substrate cleavage stops completely. The activity change is illustrated by the blue-dashed line in **(B)** and represents a guide to the eye (100 mM NaCl, 10 mM phosphate buffer, pH 6.5, 25°C). **(C)** Carbon dioxide can also be used for the stimulation of lateral pressure pulses. The slow conversion of $CO_2$ to $HCO_3^-$ and $H^+$ in water results in very long wavelengths and hence in more precise intensity measurements. In this case, too, pulse-enzyme interaction manifests in a biphasic change in activity: increasing during the expanding phase (~2fold) and inhibiting during the consecutive

condensing phase. Therefore, we conclude that the observed phenomenon is independent from the type of (acidic) excitation. Panel **(D)** depicts another guide to the eye in order to illustrate the biphasic activity change (10 mM NaCl, 1 mM phosphate buffer, pH 6.5, 25°C).

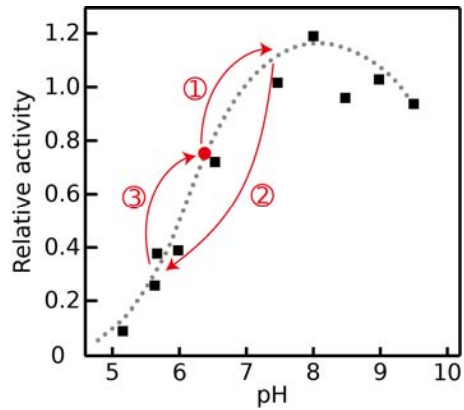

**Fig. 3.** pH dependent activity of the enzyme AChE (adapted from (30)): our measurements are performed at a pH of 6.5 in equilibrium, indicated by the red dot. The propagating pulses reversibly alter the pH at the interface. Due to the pulse induced local pH increase at the interface, we at first expect an enhanced enzyme activity (1). Consequently, the product concentration around the enzyme rises and hence also the proton concentration. Additionally, the protons, which were released during the initial expanding phase, are moving back towards the surface during the subsequent condensation of the interface back to equilibrium. This could lead to a local excess of protons and thus lower the catalytic rate of the enzyme (2). In the final step, the local pH as well as the enzyme activity reacquire their former equilibrium values and the interface adopts its initial state (3).

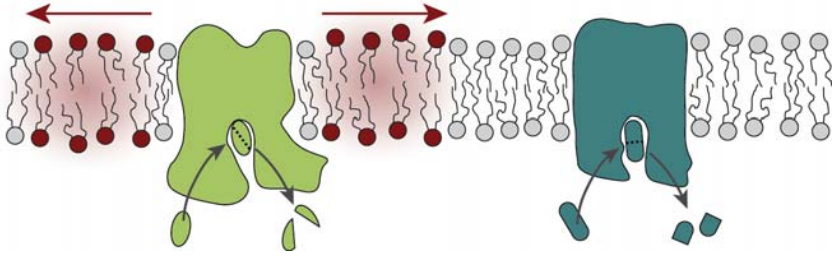

**Fig. 4.** Biological signaling by sound pulses: a perturbation, in this case induced by an enzymatic reaction, leads to a propagating signal across the interface. The excitation process can be very specific as it depends on the current state of the membrane as well as on the excitation strength. This specificity is for instance demonstrated by the pH excitation of lipid monolayers, which directly depends on the $pK_a$ of the lipids (12). A second enzyme, situated in the path of the travelling pulse, will experience a transient change of all its variables. If the pulse-induced changes are strong enough, its activity will be temporarily affected. Hence interfacial sound pulses provide a means for specific signaling in cells.

# Supplementary Information

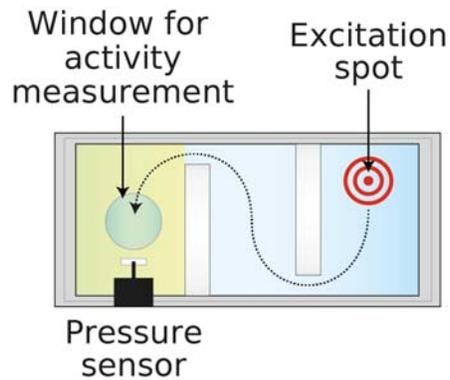

**S1:** Schematic top view of the Langmuir trough setup: Pulses are excited on a lipid monolayer via locally adding gaseous acid (HCl or $CO_2$). Two barriers prevent any unwanted gas diffusion as well as minimize convection effects in the lipid interface. A pressure sensor is used to record the mechanical component of the pulse. Acetylcholinesterase is added only in the left part of the trough, which is indicated by the yellow coloring. Its activity is monitored by the colorimetric Ellman Assay (absorption at ~410 nm).

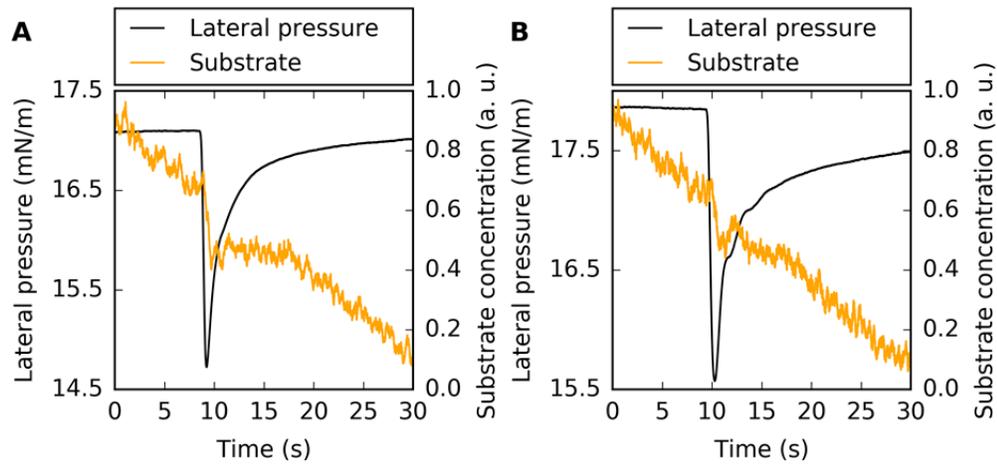

**S2:** Two further examples for the effect of HCl-induced pulses on the activity of Acetylcholinesterase: The biphasic change in activity is clearly visible in both experiments, even though, the strength varies slightly from measurement to measurement (100 mM NaCl, 10 mM phosphate buffer, 2 mM ATC, 2 mM DTNB, pH 6.5, 25°C).

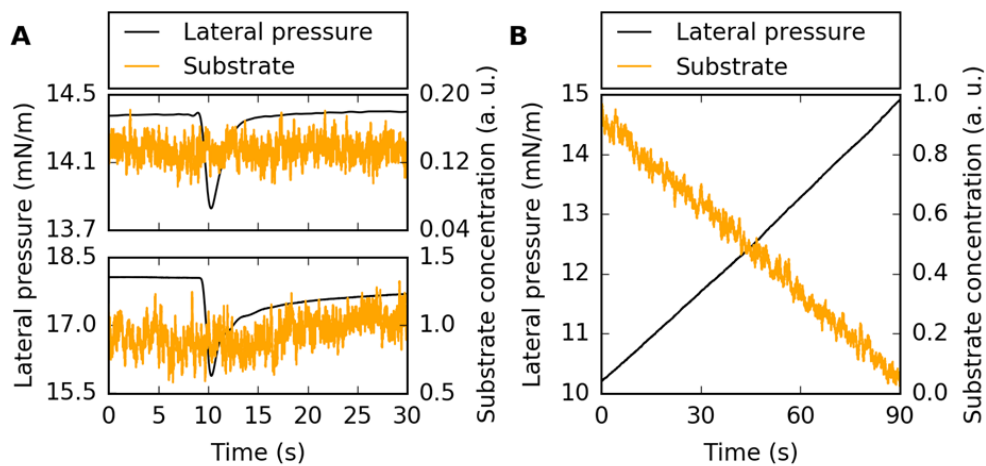

**S3:** Reference measurements: **(A)** The pulse has a negligible influence on the chromophoric substrate ATC as well as on the reaction product of the Ellman Assay. Thus any change in activity has to be attributed to the enzyme. **(B)** During isothermal compression of the monolayer in the liquid-expanded phase the intensity signal changes almost linearly, illustrating the almost constant enzyme activity in the quasi-static case (100 mM NaCl, 10 mM phosphate buffer, 2 mM ATC, 2 mM DTNB, pH 6.5, 25°C).